\documentclass[11pt]{amsart}

\numberwithin{equation}{section}

\begin{document}

\title{Hermitian extension of the four-dimensional Hooke's law}
\author{S. Antoci}
\address{Dipartimento di Fisica ``A. Volta'' and I. N. F. M.,
Via Bassi 6, Pavia, Italy}
\email{Antoci@fisav.unipv.it}
\thanks{Nuovo Cimento B, in press.}

\begin{abstract}
It has been shown recently that the classical law of elasticity,
expressed in terms of the displacement three-vector and of
the symmetric deformation three-tensor, can be extended to the
four dimensions of special and of general relativity with a
physically meaningful outcome. In fact, the resulting
stress-momentum-energy tensor can provide a unified account of
both the elastic and the inertial properties of uncharged matter.
The extension of the displacement vector to the four dimensions of
spacetime allows a further possibility. If the
real displacement four-vector $\xi_i$ is complemented with an
imaginary part $\varphi_i$, the resulting complex ``displacement''
four-vector allows for a complex, Hermitian generalisation of the
four-dimensional Hooke's law.\par
Let the complex, Hermitian ``stress-momentum-energy''
tensor density ${\bf{T}}^{ik}$ built in this way be subjected
to the ``conservation condition'' ${\bf{T}}^{ik}_{~;k}=0$.
It turns out that, while the real part of the latter equation
is able to account for the motion of electrically charged, elastic
matter, the imaginary part of the same equation can describe
the evolution of the electromagnetic field and of its sources.
The Hermitian extension of Hooke's law is performed by availing of
the postulate of  ``transposition invariance'', introduced in 1945 by
A.~Einstein for finding the nonsymmetric generalisation of his
theory of gravitation of 1915.
\end{abstract}
\maketitle

\section{Introduction}
It is well known that A. Einstein spent the last decade of his
life in the search of a non-symmetric extension of his symmetric,
Riemannian theory of 1915 \cite{ref:Einstein45},\cite{ref:Einstein55}.
The very idea of such a generalisation dates back to 1925
\cite{ref:Einstein25}, when Einstein first tried to unify the
description of gravitation and of electromagnetism through an
extension of the Riemannian geometry built from a nonsymmetric
fundamental tensor and a nonsymmetric affine connection. But
relaxing the symmetry of the geometrical objects that
enter the equations of general relativity means an
unwelcome arbitrariness in the choice of the generalisation.
In Einstein's words \cite{ref:Einstein53}:
\begin{quotation}
The introduction of non-symmetric fields meets with the
following difficulty. If $\Gamma^l_{ik}$ is a displacement field,
so is $\tilde{\Gamma}^l_{ik}~(=\Gamma^l_{ki})$. If $g_{ik}$ is a
tensor, so is $\tilde{g}_{ik}~(=g_{ki})$. This leads to a large
number of covariant formations among which it is not possible to
make a selection on the principle of relativity alone.
\end{quotation}
The way out from such an arbitrariness was envisaged by Einstein
in the requirement of ``transposition invariance'' as a
generalisation of the principle of symmetry. A tensorial expression
built with the fundamental form $g_{ik}$ and with the affine connection
$\Gamma^l_{ik}$ is said to be transposition invariant with respect
to the pair of indices, say, $p$ and $q$, if it is transformed
into itself when one simultaneously substitutes $\tilde{g}_{ik}$
for $g_{ik}$, $\tilde{\Gamma}^l_{ik}$ for $\Gamma^l_{ik}$ and
then interchanges the indices $p$ and $q$. The requirement of
transposition invariance played a key r\^ole in obtaining what
Einstein's considered the logically most satisfactory solution of
his problem, {\it i.e.} the field equations of the metric-affine
theory \cite{ref:Einstein55}:
\begin{eqnarray}
g_{ik,l}-g_{nk}\Gamma^n_{il}-g_{in}\Gamma^n_{lk}=0,\label{1.1}\\
\Gamma^s_{[is]}=0,\label{1.2}\\
R_{(ik)}=0,\label{1.3}\\
R_{[ik],l}+R_{[kl],i}+R_{[li],k}=0;\label{1.4}
\end{eqnarray}
whether a given expression be symmetric or antisymmetric with
respect to a pair of indices, say $p$ and $q$, is henceforth
denoted by enclosing the mentioned indices within respectively
round or square brackets; $R_{ik}$ means the usual Ricci
tensor
\begin{equation}\label{1.5}
R_{ik}=\Gamma^a_{ik,a}-\Gamma^a_{ia,k}-\Gamma^a_{ib}\Gamma^b_{ak}
+\Gamma^a_{ik}\Gamma^b_{ab}
\end{equation}
built with the nonsymmetric connection
$\Gamma^l_{ik}$.\par Einstein's theory of the nonsymmetric field
came in two versions, according to whether the fundamental tensor
$g_{ik}$ was chosen to be real nonsymmetric or complex Hermitian.
Much effort was done by many authors in order to grasp the
physical meaning of the proposed field equations; they looked for
solutions either by exact or by approximate methods that mimicked the
ones used in general relativity. All these efforts, however, either
implicity or explicitly allowed for singularities at the
right-hand side of one or another of the equations
(\ref{1.1})-(\ref{1.4}), while Einstein believed that, at variance with
what occurs with his theory of 1915, the new theory had to be considered
complete, and only everywhere regular solutions could disclose its physical
meaning. Therefore, by agreeing with Einstein's conviction, one can
assert that no conclusion has been drawn up to now about the validity
of the nonsymmetric theory; this is one of the unsolved problem left as
a challenge to the skillness of future mathematicians.\par
In the meantime one may well explore whether the very concept of
invariance under transposition, that Einstein found so helpful in
choosing his non Riemannian geometry, can be of heuristic
value in some down to earth instance, dealing with some well known
chapters of classical physics. Without attempting to solve
difficult equations, one can content himself with a preliminary,
modest task: seeking whether up to now unrelated physical entities,
whose mathematical representation happens to require respectively
symmetric and antisymmetric tensors, can be given an at least formally
unified description in terms of either nonsymmetric or Hermitian tensors.

\section{The four-dimensional Hooke's law as starting point}
By availing of Cartesian coordinates and of the
three-dimensional tensor formalism, Hooke's law
\cite{ref:Hooke1678} can be written as:
\begin{equation}\label{2.1}
{\Theta}^{\lambda\mu}={\frac{1}{2}}C^{\lambda\mu\rho\sigma}
(\xi_{\rho,\sigma}+\xi_{\sigma,\rho}),
\end{equation}
where ${\Theta}^{\lambda\mu}$ is the three-dimensional tensor that
defines the stresses arising in elastic matter due to its displacement,
given by the three-vector $\xi^\rho$, from a supposedly relaxed
condition, and $C^{\lambda\mu\rho\sigma}$ is the constitutive
tensor whose build depends on the material features and on the
symmetry properties of the elastic medium. It has been
shown \cite{ref:Antoci99} that this law admits of a natural
generalisation to the four-dimensions of
the Riemannian spacetime, whose metric tensor be $g_{ik}$. From a
formal standpoint, this extension is straightforward: one introduces a real
four-vector field $\xi^i$, that aims at representing some four-dimensional
displacement, and builds the four-dimensional, symmetric deformation tensor
\begin{equation}\label{2.2}
S_{ik}={1\over2}(\xi_{i;k}+\xi_{k;i}),
\end{equation}
where the semicolon indicates covariant differentiation performed
with the Christoffel symbols calculated from $g_{ik}$. A four-dimensional
stiffness tensor density ${\bf{C}}^{iklm}$
is then introduced; it will be real and symmetric with respect to both
the first and the second pair of indices, since it will be used for
producing the real symmetric stress-momentum-energy tensor density
\begin{equation}\label{2.3}
{\bf{T}}^{ik}={\bf{C}}^{iklm}S_{lm},
\end{equation}
through the generalisation of equation (\ref{2.1}) to the four dimensions
of spacetime. This generalisation happens to make physical sense,
since it allows encompassing both inertia and elasticity in a sort of
four-dimensional elasticity \cite{ref:Antoci99}. Let us consider
a coordinate system such that, at a given event, $g_{ik}$ reduce to
the form, say:
\begin{equation}\label{2.4}
g_{ik}=\eta_{ik}\equiv diag(1,1,1,-1),
\end{equation}
while the Christoffel symbols are vanishing, and the components of the
four-velocity of matter are:
\begin{equation}\label{2.5}
u^1=u^2=u^3=0,\ u^4=1.
\end{equation}
We imagine that in such a coordinate system we are able to
measure, at the chosen event, the three components of the
(supposedly small) spatial displacement of matter from its relaxed
condition, and that we adopt these three numbers as the values
taken by $\xi^\rho$ in that coordinate system, while the reading
of some clock ticking the proper time and travelling with the
medium will provide the value of the ``temporal displacement''
$\xi^4$ in the same coordinate system. By applying this procedure
to all the events of the manifold where matter is present and by
reducing the collected data to a common, arbitrary coordinate
system, we can define the vector field $\xi^i(x^k)$. From such a
field we shall require that, when matter is not subjected to
ordinary strain and is looked at in a local rest frame belonging
to the ones defined above, it will exhibit a deformation
tensor $S_{ik}$ such that its only nonzero component will be
$S_{44}=\xi_{4,4}=-1$. This requirement is met if we define the
four-velocity of matter through the equation
\begin{equation}\label{2.6}
\xi^i_{;k}u^k=u^i.
\end{equation}
The latter definition holds provided that
\begin{equation}\label{2.7}
det(\xi^i_{;k}-\delta^i_k)=0,
\end{equation}
and this shall be one equation that the field $\xi^i$ must
satisfy; in this way the number of independent components of
$\xi^i$ will be reduced to three. A four-dimensional stiffness tensor
$C^{iklm}$ endowed with physical meaning can be built as
follows. We assume that in a locally Minkowskian rest frame the
only nonvanishing components of $C^{iklm}$ are:
$C^{\lambda\nu\sigma\tau}$, with the meaning of ordinary elastic
moduli, and
\begin{equation}\label{2.8}
C^{4444}=-\rho,
\end{equation}
where $\rho$ measures the rest density of matter. We
need defining the four-dimensional stiffness tensor in an
arbitrary co-ordinate system; this task is easily accomplished
if the unstrained matter is isotropic when looked at in a locally
Minkowskian rest frame. Let us define the auxiliary
metric
\begin{equation}\label{2.9}
\gamma^{ik}=g^{ik}+u^iu^k;
\end{equation}
then the part of $C^{iklm}$ accounting for the ordinary elasticity
of the isotropic medium can be written as \cite{ref:Choquet73}
\begin{equation}\label{2.10}
C^{iklm}_{{\rm el}}=-\lambda\gamma^{ik}\gamma^{lm}
-\mu(\gamma^{il}\gamma^{km}+\gamma^{im}\gamma^{kl}),
\end{equation}
where $\lambda$ and $\mu$ are assumed to be constants. The part of
$C^{iklm}$ that accounts for the inertia of matter shall read
instead
\begin{equation}\label{2.11}
C^{iklm}_{{\rm in}}=-\rho u^iu^ku^lu^m.
\end{equation}
The elastic part $T^{ik}_{{\rm el}}$ of the energy tensor is orthogonal
to the four-velocity, as it should be \cite{ref:Cattaneo71}; thanks to
equation (\ref{2.6}) it reduces to
\begin{eqnarray}\label{2.12}
T^{ik}_{{\rm el}}=C^{iklm}_{{\rm el}}S_{lm}
=-\lambda(g^{ik}+u^iu^k)(\xi^m_{;m}-1)\nonumber\\
-\mu[\xi^{i;k}+\xi^{k;i}+u_l(u^i\xi^{l;k}+u^k\xi^{l;i})],
\end{eqnarray}
while, again thanks to equation (\ref{2.6}), the inertial part of
the energy tensor proves to be effectively so, since
\begin{equation}\label{2.13}
T^{ik}_{{\rm in}}=C^{iklm}_{{\rm in}}S_{lm} =\rho u^iu^k.
\end{equation}
The energy tensor defined by summing the contributions
(\ref{2.12}) and (\ref{2.13}) encompasses both the inertial and
the elastic energy tensor of an isotropic medium; when the
macroscopic electromagnetic field is vanishing it should represent
the overall energy tensor, whose covariant divergence must vanish
according to Einstein's equations \cite{ref:Hilbert15},
\cite{ref:Klein17}:
\begin{equation}\label{2.14}
T^{ik}_{;k}=0.
\end{equation}
Imposing the latter condition allows one to write the equations of
motion for isotropic matter subjected only to elastic strain
\cite{ref:Cattaneo71}. We show this outcome in the limiting case when
the metric is everywhere flat and the four-velocity of matter is
such that $u^\rho$ can be dealt with as a first order
infinitesimal quantity, while $u^4$ differs from unity at most for
a second order infinitesimal quantity. Also the spatial components
$\xi^\rho$ of the displacement vector and their derivatives are
supposed to be infinitesimal at first order. An easy calculation
\cite{ref:Antoci99} then shows that equation (\ref{2.7}) is satisfied
to the required first order, and that equations (\ref{2.14})
reduce to the three equations of motion:
\begin{equation}\label{2.15}
\rho\xi^\nu_{,4,4}=\lambda\xi^{\rho,\nu}_{~,\rho}
+\mu(\xi^{\nu,\rho}+\xi^{\rho,\nu})_{,\rho},
\end{equation}
and to the conservation equation
\begin{equation}\label{2.16}
\{\rho u^4u^k\}_{,k}=0,
\end{equation}
{\it i.e.}, to the required order, they come to coincide with the
well known equations of the classical theory of elasticity for an
isotropic medium.

\section{Hermitian extension of the four-dimensional Hooke's law}
The extension of Hooke's law outlined in the previous Section has
been a fruitful move, since it has allowed a unified account of
inertia and of elasticity. But there is another aspect that deserves
attention: having enlarged Hooke's law to the four dimensions of
spacetime opens the way to this new question: is it useful, {\it i.e.},
does it make physical sense to look for either a nonsymmetric or a Hermitian
extension of the four-dimensional Hooke's law?\par
Let us try the Hermitian version, and see how it can be formulated
in a Riemannian spacetime whose metric, given a priori, is the real,
symmetric tensor $g_{ik}$. We introduce a complex ``displacement''
four-vector
\begin{equation}\label{3.1}
\omega^i\equiv\xi^i+i\varphi^i;
\end{equation}
the real four-vectors $\xi^i$ and $\varphi^i$ enter respectively the
real and the imaginary part of $\omega^i$. By closely following
the pattern of the real case, in lieu of (\ref{2.2})
we introduce the Hermitian ``deformation'' tensor
\begin{equation}\label{3.2}
S_{ik}={1\over2}(\omega^*_{i;k}+\omega_{k;i});
\end{equation}
the asterisk is henceforth used to denote complex conjugation. $S_{ik}$
splits into a real, symmetric part $S_{(ik)}$, that
can still be interpreted as a genuine deformation tensor:
\begin{equation}\label{3.3}
S_{(ik)}={1\over2}(\xi_{i;k}+\xi_{k;i})
\end{equation}
and into an antisymmetric, purely imaginary contribution:
\begin{equation}\label{3.4}
S_{[ik]}={i\over2}(\varphi_{k,i}-\varphi_{i,k})
\end{equation}
that immediately reminds us of Helmholtz' seminal attempt at
producing a hydrodinamic {\it simile} of magnetism
\cite{ref:Whittaker51}, and of the subsequent extension
of the idea to the four dimensions of the theory of relativity.
We assume that $\varphi_i$ shall play the r\^ole of the electromagnetic
four-potential and, starting from this apparently promising beginning,
we try to figure out what the idea of a Hermitian generalisation of
Hooke's law can entail from a physical standpoint.\par
Instead of the stiffness tensor density of equation (\ref{2.3}),
that is symmetric in both the first and the second pair of indices,
one shall confront a complex ``stiffness'' tensor density ${\bf{C}}^{iklm}$,
that will be Hermitian with respect to both the first and the
second pair:
\begin{equation}\label{3.5}
{\bf{C}}^{kilm}={\bf{C}}^{*iklm}={\bf{C}}^{ikml}.
\end{equation}
This stiffness tensor density shall be contracted with the
Hermitian deformation tensor $S_{ik}$ to produce the Hermitian
tensor density
\begin{equation}\label{3.6}
{\bf{T}}^{ik}={\bf{C}}^{iklm}S_{lm}.
\end{equation}
One calls for now ${\bf{T}}^{ik}$ the Hermitian stress-momentum-energy
tensor density, and asks that it fulfil the generalised
``conservation'' equation
\begin{equation}\label{3.7}
{\bf{T}}^{ik}_{~;k}=0,
\end{equation}
which of course entails severally:
\begin{equation}\label{3.8}
{\bf{T}}^{(ik)}_{~;k}=0,
\end{equation}
and
\begin{equation}\label{3.9}
{\bf{T}}^{[ik]}_{~,k}=0.
\end{equation}
A survey of the tensorial parts into which (\ref{3.6}) can be split
and expanded in keeping with the symmetry properties is now useful.
The real, symmetric part of (\ref{3.6}) can be written as:
\begin{equation}\label{3.10}
{\bf{T}}^{(ik)}={\bf{C}}^{(ik)(lm)}S_{(lm)}+{\bf{C}}^{(ik)[lm]}S_{[lm]}.
\end{equation}
The interpretation of the first term at the right-hand side has been
already provided, since $S_{(lm)}$ is just the four-dimensional, symmetric
deformation tensor of Section 2; as recalled there, if $S_{(lm)}$ is
contracted with the appropriate ${\bf{C}}^{(ik)(lm)}$, it can
produce the material part of the symmetric energy tensor density, that
accounts for both the inertia and the elasticity of matter. Let us call it
${\bf{T}}^{(ik)}_{\rm m}$.\par
Due to the presence of $S_{[lm]}$, the second term at the right-hand
side of (\ref{3.10}) awaits an electromagnetic interpretation. One
tentatively sets
\begin{equation}\label{3.11}
F_{ik}\equiv-2iS_{[ik]},
\end{equation}
where $F_{ik}$ is the antisymmetric tensor used in
electromagnetism, according to a long established convention,
to encompass both the electric field $\vec{E}$ and the magnetic
induction $\vec{B}$.
A symmetric energy tensor density ${\bf{T}}^{(ik)}_{\rm em}$ for
the electromagnetic field in matter, like {\it e.g.} Abraham's tensor
density \cite{ref:Abraham09}, \cite{ref:Gordon23} can obviously be
recast in the form\footnote{It may be objected that this
way of writing the electromagnetic energy tensor is artificial,
since the four-potential $\varphi_i$ enters not only $S_{[lm]}$,
but also the ``stiffness'' factor ${\bf{C}}^{(ik)[lm]}$. However
the previous Section showed that a similar occurrence already
happens with the displacement vector $\xi_i$ in
${\bf{T}}^{(ik)}_{\rm m}$ as soon as one abandons the
non-relativistic regime.}
\begin{equation}\label{3.12}
{\bf{T}}^{(ik)}_{\rm em}={\bf{C}}^{(ik)[lm]}S_{[lm]};
\end{equation}
therefore the overall ${\bf{T}}^{(ik)}$ is apt to summarise the
energy tensor density of elastic matter and of the electromagnetic
field. Equating to zero the four-divergence of
this density, as done in (\ref{3.8}), would produce four equations
for the motion of possibly charged elastic matter under the
influence of an electromagnetic field.\par
We consider now the imaginary, antisymmetric part of
${\bf{T}}^{ik}$, that can be written as
\begin{equation}\label{3.13}
{\bf{T}}^{[ik]}={\bf{C}}^{[ik](lm)}S_{(lm)}+{\bf{C}}^{[ik][lm]}S_{[lm]}.
\end{equation}
The second term at the right-hand side of this equation naturally
fits the physical picture already emerged from the analysis of
${\bf{T}}^{(ik)}$. In fact, since $-2iS_{[ik]}$ has already been
identified in (\ref{3.11}) as representing the electric field
$\vec{E}$ and the magnetic induction $\vec{B}$, one is led to pose
\begin{equation}\label{3.14}
{\bf H}^{ik}\equiv-i{\bf{T}}^{[ik]}_{\rm f}
\equiv-i{\bf{C}}^{[ik][lm]}S_{[lm]},
\end{equation}
{\it i.e.} to read off the second term at the right-hand side
of (\ref{3.13}) the so called constitutive equation of
electromagnetism \cite{ref:Post62}. This equation defines
the electric displacement $\vec{D}$ and of the magnetic field
$\vec{H}$, summarised by the four-tensor density ${\bf H}^{ik}$, in
terms of $\vec{E}$, $\vec{B}$, and of whatever
fields may enter ${\bf{C}}^{[ik][lm]}$. For definiteness,
let us remind an example of the constitutive equation, valid when matter
is homogeneous and isotropic in its local rest frame \cite{ref:Gordon23}:
\begin{equation}\label{3.15}
\mu{H^{ik}}=\big[g^{il}-(\epsilon\mu-1)u^{i}u^{l}\big]
\big[g^{km}-(\epsilon\mu-1)u^{k}u^{m}\big]F_{lm};
\end{equation}
here $\epsilon$ is the dielectric constant and $\mu$ is the
magnetic permeability. This constitutive equation has just the
form (\ref{3.14}).\par
If only the second term ${\bf{C}}^{[ik][lm]}S_{[lm]}$ where
present at the right-hand side of (\ref{3.13}), the imaginary
part (\ref{3.9}) of the ``conservation'' equation would entail
${\bf H}^{ik}_{~,k}=0$, {\it i.e.} the electromagnetic field would
be sourceless, and our description of matter would be defective.
The Hermitian extension of the four-dimensional Hooke's
law however provides a solution to this problem through
the first term at the right-hand side of (\ref{3.13}).
Let us define:
\begin{equation}\label{3.16}
{\bf P}^{ik}\equiv i{\bf{T}}^{[ik]}_{\rm ch}
\equiv i{\bf{C}}^{[ik](lm)}S_{(lm)},
~~{\bf j}^i\equiv {\bf P}^{ik}_{~,k},
\end{equation}
where ${\bf j}^i$ is a conserved quantity, since it is
the divergence of an antisymmetric tensor density:
\begin{equation}\label{3.17}
{\bf j}^i_{~,i}=0.
\end{equation}
Equation (\ref{3.9}) can be rewritten as
\begin{equation}\label{3.18}
{\bf{T}}^{[ik]}_{{\rm ch}~,k}+{\bf{T}}^{[ik]}_{{\rm f }~,k}=0,
\end{equation}
and, due to the definitions (\ref{3.14}) and (\ref{3.16}), it
entails:
\begin{equation}\label{3.19}
{\bf H}^{ik}_{~,k}={\bf j}^i,
\end{equation}
{\it i.e.} the validity of the inhomogeneous Maxwell's equation.
\section{Charge, like matter, exists since time elapses}
We have completed the attribution of a tentative physical meaning
to the four tensorial terms into which the Hermitian
${\bf{T}}^{ik}$ can be split according to the symmetry properties.
Three of them, namely the material contribution
${\bf{T}}^{(ik)}_{\rm m}$, the electromagnetic energy tensor
${\bf{T}}^{(ik)}_{\rm em}$ and the term that provides for the constitutive
equation of electromagnetism, are entities known since a long
time. Their mathematical form and their physical meaning have been carefully
investigated by generations of scholars. The very existence of the fourth one,
${\bf{T}}^{[ik]}_{\rm ch}$, is predicted by the Hermitian
extension of the four-dimensional Hooke's law.
It is certainly welcome from a physical standpoint. Its very build,
however, would be a surprise, had we not already met in
Section 2 with an occurrence that constitutes its symmetric counterpart.
We have seen there that the extension of Hooke's law to the
four dimensions of spacetime allows one to
account, besides the ordinary elasticity, also for the very existence
of inertial matter. This happens thanks to the completion of the
displacement vector with a fourth component which, as shown in
Section 2, has the meaning of displacement in time. In the same way
we must expect that the term ${\bf{C}}^{[ik](lm)}S_{(lm)}$
shall account, besides the phenomenon of ordinary piezoelectricity,
also for the very existence of the electric charge and current in
unstrained matter. Through the generalised deformation tensor
$S_{(lm)}$ the relativistic and Hermitian extension of Hooke's law
relates the presence of both matter and charge to the lapsing of time.
\section{Perspectives}
One still knows nothing about the explicit expression of
${\bf{C}}^{[ik](lm)}$, and it is fully premature to address here the
enormous task of providing models that best suit the
manifold properties displayed by electricity in macroscopic matter. Of
course one shall proceed in this undertaking by adhering to the pattern
already followed with the other three terms that compose ${\bf{C}}^{iklm}$,
{\it i.e.} one shall try to build this fourth term by availing only of
the two four-vectors $\xi_i$, $\varphi_i$, of the scalar density
$\rho$ and of the metric tensor $g_{ik}$ This is possible, since (\ref{2.6})
allows expressing the four-velocity $u^i$ through $\xi_i$ and
$g_{ik}$. The postulate of Hermitian symmetry
will restrict the choice of the possible forms.
In the present formulation $g_{ik}$ is an entity prescribed from the outside,
and we know from the condition (\ref{2.7}) that $\xi_i$ has only three
independent components. Pending a detailed scrutiny of the Cauchy problem,
we can notice that the complex equation ${\bf{T}}^{ik}_{~;k}=0$ imposes
eight conditions on the eight independent variables of our problem.
The model of electrified, elastic matter provided by the Hermitian
extension of the four-dimensional Hooke's law has therefore a chance to
stand up as a complete model, in which the values of all the
quantities accounting for the physical behaviour of both matter and the
electromagnetic field can be determined at least in principle by solving,
with given initial conditions, the equations of motion stemming from
the Hermitian ``conservation equation''.

\bibliographystyle{amsplain}

\end{document}